\pdfoutput=1
\documentclass{article}

\usepackage{microtype}
\usepackage{graphicx}
\usepackage{subfigure}
\usepackage{booktabs} % for professional tables

\usepackage{hyperref}
\usepackage{url}
\usepackage{graphicx}
\usepackage{makecell}
\usepackage{multirow}
\usepackage{enumitem}
\usepackage{bbm}
\usepackage{color,xcolor}
\usepackage{booktabs}
\usepackage{lineno}
\usepackage{array}
\usepackage{ulem}
\usepackage{makecell}
\usepackage{multicol}
\usepackage{amsmath,amsthm,amsfonts,amssymb,bm}
\usepackage{stfloats}
\usepackage[font=small,labelfont=bf]{caption}
\usepackage{tabulary}
\usepackage{algorithm}
\usepackage{algorithmicx}
\usepackage{algpseudocode}
\newcolumntype{C}[1]{>{\centering\arraybackslash}p{#1}}
\newcolumntype{L}[1]{>{\arraybackslash}p{#1}}
\DeclareMathOperator*{\argmax}{arg\,max}

\newcommand\NoThen{\renewcommand\algorithmicthen{}}

\newcommand\NoDo{\renewcommand\algorithmicdo{}}

\algnewcommand{\algorithmicendif}{}
\algnewcommand{\algorithmicforend}{}

\algtext*{EndIf}
\algtext*{EndFor}

\usepackage[preprint]{neurips_2023}

\author{Zhangyang Gao$^{\dagger}$, Xingran Chen$^{\dagger}$, Cheng Tan$^{\dagger}$, Stan Z. Li$^{*}$\\
AI Lab, Research Center for Industries of the Future, Westlake University \\
\texttt{\{gaozhangyang, tancheng,Stan.ZQ.Li\}@westlake.edu.cn}\\
\texttt{chenxran@umich.edu}\\
\thanks{$^{\dagger}$Equal Contribution, $^{*}$Corresponding Author.}
}
\makeatletter
\def\thanks#1{\protected@xdef\@thanks{\@thanks
        \protect\footnotetext{#1}}}
\makeatother

\usepackage[utf8]{inputenc} % allow utf-8 input
\usepackage[T1]{fontenc}    % use 8-bit T1 fonts
\usepackage{hyperref}       % hyperlinks
\usepackage{url}            % simple URL typesetting
\usepackage{booktabs}       % professional-quality tables
\usepackage{amsfonts}       % blackboard math symbols
\usepackage{nicefrac}       % compact symbols for 1/2, etc.
\usepackage{microtype}      % microtypography
\usepackage{xcolor}         % colors
\usepackage{wrapfig}

\title{MotifRetro: Exploring the  Combinability-Consistency Trade-offs in retrosynthesis via Dynamic Motif Editing}

\begin{document}

\maketitle

\begin{abstract}
  Is there a unified framework for graph-based retrosynthesis prediction? Through analysis of full-, semi-, and non-template retrosynthesis methods, we discovered that they strive to strike an optimal balance between combinability and consistency: \textit{Should atoms be combined as motifs to simplify the molecular editing process, or should motifs be broken down into atoms to reduce the vocabulary and improve predictive consistency?}
  Recent works have studied several specific cases, while none of them explores different combinability-consistency trade-offs. Therefore, we propose MotifRetro, a dynamic motif editing framework for retrosynthesis prediction that can explore the entire trade-off space and unify graph-based models. MotifRetro comprises two components: RetroBPE, which controls the combinability-consistency trade-off, and a motif editing model, where we introduce a novel LG-EGAT module to dynamiclly add motifs to the molecule. We conduct extensive experiments on USPTO-50K to explore how the trade-off affects the model performance and finally achieve state-of-the-art performance.
\end{abstract}

\section{Introduction}

\begin{figure}[h]
  \centering
  \includegraphics[width=5.5in]{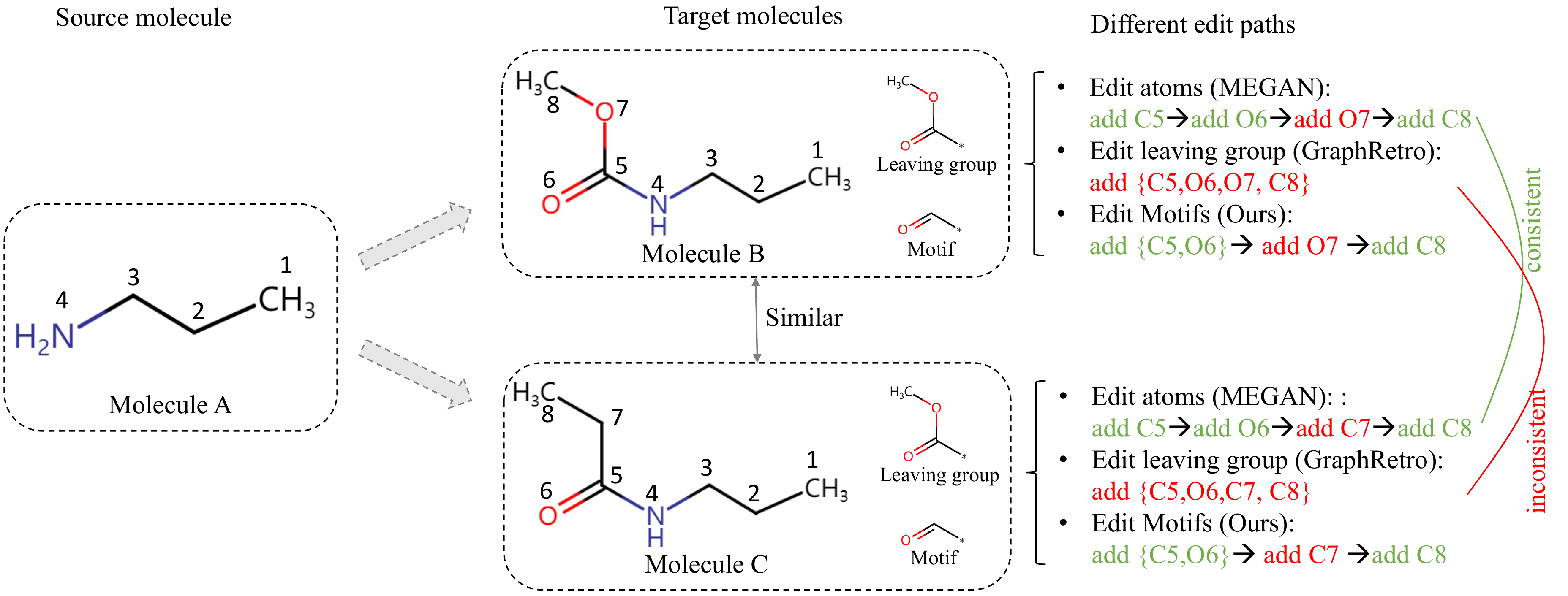}
  \caption{Consistency and combinability. We illustrate editing paths from molecule A to molecule B and molecule A to molecule C, where combinability refers to the ability to combine atoms into motifs, and consistency describes the similarity of editing paths for generating molecule B and C. There are two extreme cases: (1) Extreme consistency. If the basic editing units are atoms, the edit paths are fairly consistent, with three consistent edit steps marked in green and one inconsistent edit step marked in red. (2) Extreme composability. When leaving groups serve as the basic editing units, the resulting editing paths become inconsistent, with only one inconsistent edit step marked in red. However, there is an additional benefit: the edit path length is shortened to 1, thus reducing the search space. Previous methods, such as MEGAN \citep{sacha2021molecule} and GraphRetro \citep{somnath2021learning}, either edit atoms or leaving groups. They do not consider decomposing leaving groups into motifs, which can provide both high consistency and reduced search space, leading to a better trade-off between composability and consistency.}

  \label{fig:combinability_consistency}
\end{figure}

Single-step retrosynthesis prediction, a fundamental chemical problem that predicts the inverse chemical reaction, plays a critical role in synthesis planning and drug discovery \citep{corey1969computer,corey1991logic}. However, due to the complex molecular structure changes, multiple theoretically correct synthetic paths, and vast search space, the problem is highly challenging and requires considerable expertise and experience. Recently, deep learning methods \cite{shi2020graph,NEURIPS2020_819f46e5,coley2017computer,segler2017neural,zheng2019predicting,chen2019learning,somnath2021learning,sacha2021molecule,seidl2021modern,sun2020energy,wan2022retroformer,gao2022semiretro,liu2022mars} have offered new opportunities in retrosynthesis prediction by enabling the discovery of new knowledge from big data beyond human experts and potentially providing a more effective way to optimize chemical production processes. Combining expert experience with deep learning is a promising direction for retrosynthesis prediction, and some recent studies \cite{gao2022semiretro,liu2022mars, dai2019retrosynthesis, somnath2021learning} have followed this approach by introducing structural motifs into retrosynthesis, which is essentially a molecular editing problem. As shown in Figure~ \ref{fig:combinability_consistency}, introducing motifs can simplify the molecular editing process, but it can also increase the vocabulary size and decrease predictive consistency. The reader can view Figure~\ref{fig:combinability_consistency} to understand the concept of consistency \footnote{We recommend the reader to view Figure~\ref{fig:combinability_consistency} for the intuition of the concept of consistency and combinability. Formal definitions are described in Sec.\ref{sec:retrobpe}.}.

% 研究现状, 挑战, 标签构造的差异
Current machine learning-based retrosynthesis work can be classified based on the degree of use of structural priors: full-template (FT), semi-template (ST), and non-template (NT). Full-template methods \cite{segler2017neural, coley2017computer, dai2019retrosynthesis, chen2021deep} predict reaction templates directly, where the templates encode all structural changes in the reaction. Recent semi-template methods \cite{somnath2021learning, gao2022semiretro, liu2022mars} simplify full templates to simpler motif components to achieve better consistency and higher accuracy. However, the existing templates can confine the inference process, raising concerns about the generalization ability of template-based models. To address this issue, many non-template approaches have been proposed. For example, transformer models \cite{zheng2019predicting, chen2019learning,NEURIPS2020_819f46e5,wan2022retroformer} treat molecule as SMILES sequences and model the retrosynthesis prediction as a sequence-to-sequence problem; graph editing models~\cite{shi2020graph, sacha2021molecule} manipulate molecules at the atom level.

From full-template, semi-template to non-template, there is a trade-off between the combinability and consistency. Relying solely on templates can limit generalizability due to minor structural modifications that can lead to significant changes in the template, causing confusion in the model's predictions. For instance, although molecule B and molecule C in Figure~\ref{fig:combinability_consistency} differ by only one atom, their leaving groups are considered different templates. Conversely, getting rid of templates would make the model less constrained, leading to poorer performance. Unfortunately, there have been no studies exploring how to achieve a better  trade-off between combinability and consistency. To the best of our knowledge, this is the first work investigating the trade-off between combinability and consistency. Several key challenges exist in studying the trade-off. Firstly, there is no controllable algorithm for decomposing complex reaction templates into simpler motifs for human purposes. Secondly, reducing the single-step inverse synthesis problem to a multi-step graph editing problem is challenging, especially with the operation of adding motifs, which requires significant effort in correcting and satisfying chemical constraints. Overall, developing a unified framework for retrosynthesis prediction while considering the combinability-consistency trade-off is a non-trival task that involves significant effort in data processing and algorithm design.

We propose MotifRetro, an end-to-end motif-based graph editing model for retrosynthesis prediction. MotifRetro employs a dynamic graph editing process consisting of three phases: bond breaking, motif addition, and bond formation. The input molecule (product) is first converted to synthons by breaking bonds, then repeatedly adding new structural motifs to the existing molecule, and finally forming new bonds to support the ring-opening reaction and getting output molecules (reactants). To achieve diverse combinability-consistency trade-offs, we propose a novel RetroBPE algorithm that decomposes each leaving group added to the input molecule as a motif tree, where each leaf is a simple structural motif. In this way, we replace the low-frequency leaving groups with high-frequency motifs to achieve better consistency and generalizability which is shown in their top-10 accuracy.  Additionally, the structural knowledge of motifs improves top-1 and top-3 accuracy by reducing the search space compared to template-free models. The controllable combinability and consistency of the motif tree allow us to explore a broader range of trade-offs, and MotifRetro can unify several existing methods \cite{liu2022mars,gao2022semiretro,somnath2021learning,shi2020graph,sacha2021molecule} as specific cases under certain trade-offs. In summary, our contributions include:

\begin{enumerate}
  \item We introduce the concept of motif tree and propose a novel RetroBPE methods for controling the combinability-consistency trade-offs. 
  \item We propose an LG-EGAT layer to build the dynamic motif editing model, which converts the single-step retrosynthesis prediction to a multi-step motif editing problem.
  \item We explore the full trade-off space on USPTO-50K and show how the trade-off affects the model performance.
  \item The proposed MotifRetro framework unifies many previous graph-based retrosynthesis algorithms and achieves state-of-the-art retrosynthesis performance.
\end{enumerate}
\section{Related Work}
\subsection{Retrosynthesis Analysis}
\paragraph{Problem definition} In this work, we denote the molecular state as $\mathcal{G}(A,X,E)$, where $A$, $X$, and $E$ are the adjacency matrix, atom features, and bond features, respectively. Note that $\mathcal{G}$ may contain more than one molecule, including multiple molecules or multiple molecular fragments. Retrosynthesis aims to infer a set of reactants ($\mathcal{G}^{(T)}$) that can generate the product molecule ($\mathcal{G}^{(0)}$) through chemical reaction. Formally, this is achieved by learning a mapping function $f_{\theta}$.:
\begin{align}
    \label{eq:retrosynthesis}
    f_{\theta}: \mathcal{G}^{(0)} \mapsto \mathcal{G}^{(T)}
\end{align}
Predicting $\mathcal{G}^{(T)}$ directly from $\mathcal{G}^{(0)}$ in one step is typically difficult, therefore we decompose the task into multi-step simple tasks: $\mathcal{G}^{(0)} \rightarrow \mathcal{G}^{(1)} \rightarrow \mathcal{G}^{(2)} \rightarrow \cdots \rightarrow \mathcal{G}^{(T)}$. At each step, the model needs to edit the input molecules toward the target molecule until the final state is exactly the same as the reactant. Finally, the objective is to obtain a single-step graph editing model:
\begin{align}
    \label{eq:retrosynthesis}
    f_{\theta}: \mathcal{G}^{(t)}, t \mapsto \mathcal{G}^{(t+1)}
\end{align}

\begin{figure}[t]
    \centering
    \includegraphics[width=5.5in]{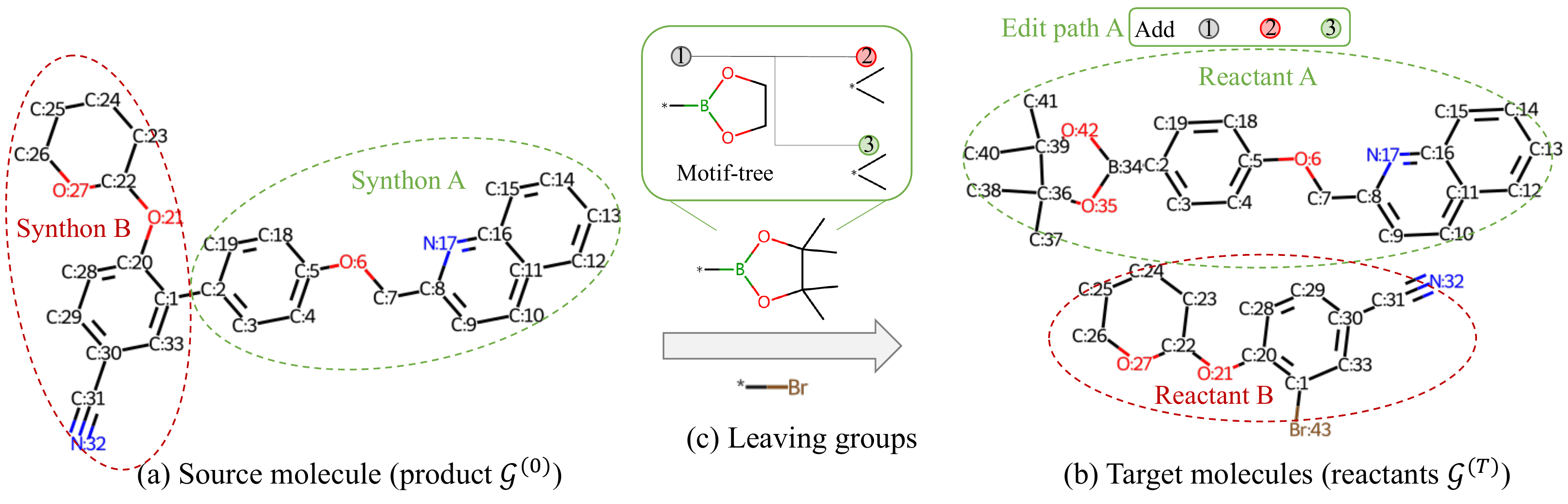}
    \caption{Problem definition. The objective is to generate reactant molecules $\mathcal{G}^{(T)}$ from the input product molecule $\mathcal{G}^{(0)}$. In this example, there are two reactants A and B that need to be predicted. Taking the editing process from synthon A to reactant A as an example, the model gradually adds 1, 2, and 3 motifs. }
    \label{fig:problem_def}
\end{figure}

\paragraph{Full template Method} 
These approaches \citep{heid2021influence, wan2021neuraltpl, chen2021deep, coley2017computer, dai2019retrosynthesis} leverage existing reaction corpora to construct a template knowledge base, which divides retrosynthesis analysis into two steps. The initial step involves the creation of a database grounded in reaction rules. Prior works such as Satoh et al. \citep{satoh1995sophia}, Hartenfeller et al. \citep{hartenfeller2011collection}, and Szymkuc et al. \citep{szymkuc2016computer} have meticulously crafted reaction rules that guide the prediction process in retrosynthetic analysis. In contrast, studies like those by Coley et al. \citep{coley2017prediction} and Law et al. \citep{law2009route} have focused on developing algorithms capable of automatically extracting reaction rules from the comprehensive reactions knowledge base. The second step revolves around matching the template to the desired end product and extrapolating the necessary reactants accordingly. The most straightforward approach to achieving this is through sub-graph matching, a method that, while effective, is computationally costly, particularly when applied to large-scale scenarios. To circumvent this issue, Coley et al. \cite{coley2017computer} suggested treating the existing set of reactions as the knowledge base and retrieving reactions for reference based on molecular similarity. Moreover, Segler et al. \cite{segler2017neural} and Baylon et al. \cite{baylon2019enhancing} incorporated a neural network to encode target molecules and learn the conditional distribution of rules given the product. Dai et al. \cite{dai2019retrosynthesis} expanded upon this work to further learn the joint conditional distribution of reaction rules and reactants given products. Due to the inherent reliance on templates, the generalizability of template-Based algorithms are also confined by these pre-made templates.

\paragraph{ Non-template Method} Unlike template-based methodologies, non-template methods directly generate the reactant molecules.  Sequential models, such as Transformer \cite{vaswani2017attention,wang2021retroprime,zheng2019predicting, chen2019learning,NEURIPS2020_819f46e5,wan2022retroformer, tu2021permutation} and LSTM \cite{hochreiter1997long}, have been used to solve retrosynthesis problems. These techniques typically operate based on SMILES representations of molecules. For instance, Graph2SMILES \citep{tu2021permutation} combines the graph encoder and sequence encoder to produce a permutation invariant generator. This approach performs well in predicting reactants that are structurally diverse and complex. RetroPrime \citep{wang2021retroprime} integrates chemists' retrosynthetic strategy into the Transformer model, making it more interpretable and easier for chemists to understand the underlying reasoning behind the model's predictions. Additionally, RetroFormer \citep{wan2022retroformer} proposes a transformer architecture that does not rely on any cheminformatics tools for retrosynthesis prediction, making it more efficient and scalable. 
Furthermore, graph-based methods such as G2G \citep{shi2020graph} and MEGAN \citep{sacha2021molecule} can also operate in the sense of non-template by editing atoms one-by-one.

\paragraph{Semi-template Method} These approaches combine the advantage of both the generative models and additional chemical knowledge. Recent studies fall in this catogory formulates retrosynthesis as a two-step process: (1) reaction center identification; (2) synthon completion. The center identification detects the reaction centers and convert the product into multiple intermediate molecules, called synthons. In synthon completion, RetroXbert\cite{NEURIPS2020_819f46e5} transform it back to a seq2seq problem to predict SMILES;  GraphRetro \cite{somnath2020learning} founds that the leaving groups in the training set have high coverage, therefore they transform the problem as a leaving group classification; SemiRetro \cite{gao2022semiretro} decompose the full reaction template into simpler ones attached to each synthons to reduce the template space. In general, semi-template methods offer more scalability than full-template ones, but they often perform poorly in terms of top-1 accuracy due to a lack of structural knowledge.  Comparably, MotifRetro can be considered either semi-template or non-template, depending on the degree of using structural motifs (combinability). This unique feature provides much-needed flexibility, allowing for achieve a good balance between template-based and template-free methods.

\paragraph{Subgraph Mining}
Existing works applying subgraph-level generation mainly focus on molecule generation. \cite{jin2018junction, jin2020hierarchical} leverage handcrafted breaking rules to decompose molecules into subgraphs and generate the molecule in subgraph level; \cite{yang2021hit} utilizes the existing chemical fragment library  with reinforcement learning in generation; \cite{kong2022molecule, yan2002gspan, nijssen2004quickstart, inokuchi2000apriori} learn a subgraph vocabulary based on the given training corpus. In comparison, this paper proposes RetroBPE, which mines the motif-tree for leaving groups used in retrosynthesis analysis. The motif-tree is characterized by (1) the root node of the tree being at the binding position of the leaving group to the synthons, (2) the ring being treated as the same basic unit as the atom, with no broken ring operation, and (3) each leaf node being connected to the tree through a shared atom.

\section{Method}
\subsection{Overall Framework}
\begin{figure}[H]
    \centering
    \includegraphics[width=5.5in]{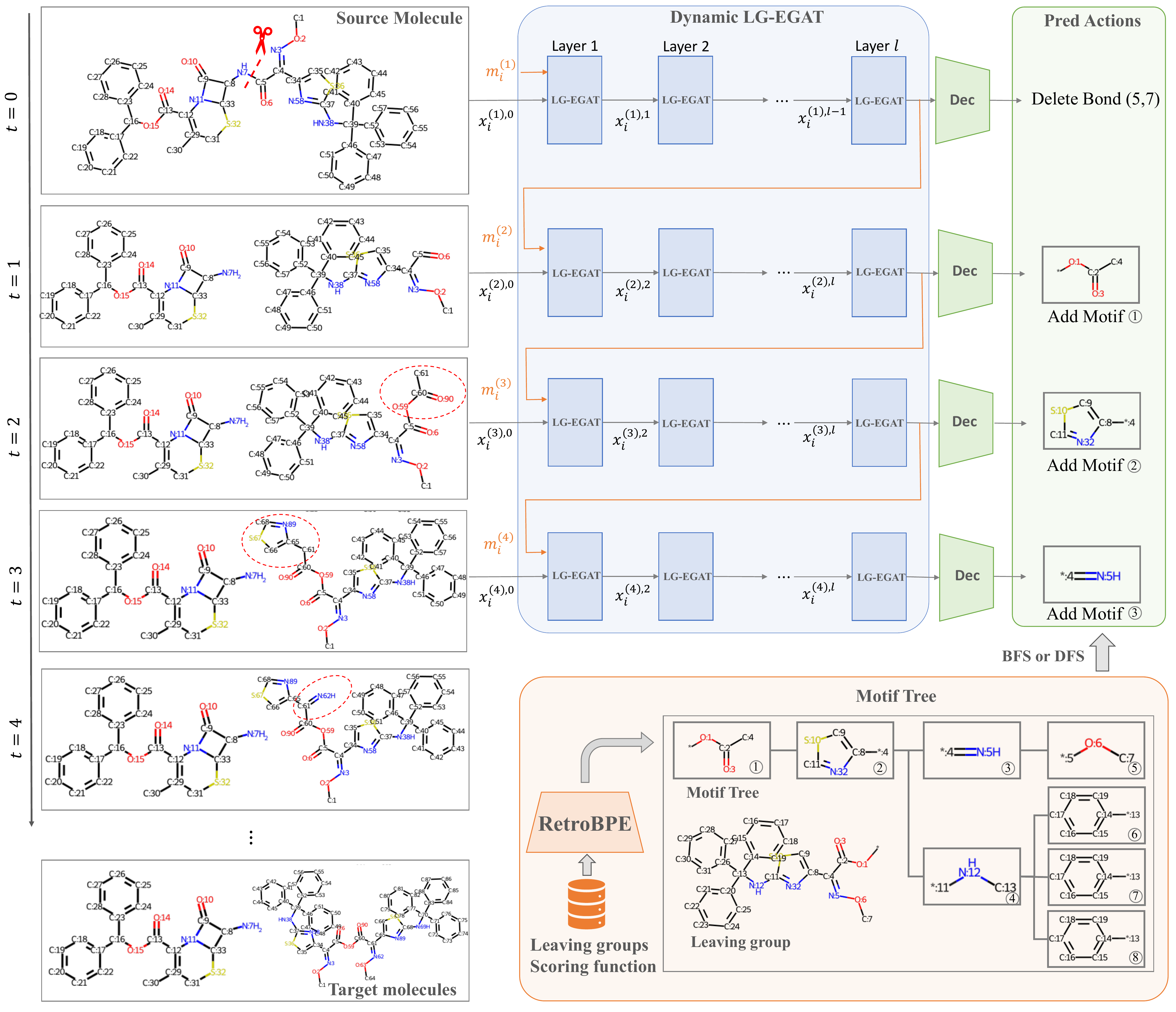}
    \caption{The overall framework. MotifRetro comprises two main components: RetroBPE, which controls the trade-off between combinability and consistency to construct motif-trees, and a motif editing model. We introduce a novel LG-EGAT module in the motif editing model to dynamically add motifs to the current molecule. At each timestamp, the model predicts an action and applies it to the molecule to generate a new molecule for the next prediction until the "Stop" action is reached.}
    \label{fig:overall_framework}
\end{figure}

We propose MotifRetro, which represents leaving groups as the motif tree, to explore the search space with various combinability-consistency trade-off by using a controllable motif extracting algorithm, called RetroBPE. As shown in Figure~\ref{fig:overall_framework}, MotifRetro learns the dynamic graph editing process for retrosynthesis prediction. By controlling the combinability-consistency trade-offs, MotifRetro offers the opportunity of unifying different graph-based retrosynthesis algorithms. In extreme cases, each leaf of the motif-tree is an atom when we emphasize the consistency like MEGAN \cite{sacha2021molecule} and G2G \cite{shi2020graph}, or the motif-tree only contains one leaf that represents the leaving group when we emphasize the combinability like GraphRetro \cite{somnath2021learning} and SemiRetro \cite{gao2022semiretro}.

\subsection{Motif Tree \& RetroBPE}
\label{sec:retrobpe}
Previous research has introduced chemical insights into decomposing the source molecule (product) into separate synthons, each of which can be translated into target molecules (reactants) by adding leaving groups. The approach for adding leaving groups leads to major algorithmic differences: (1) full-template methods predict the reaction templates at once, emphasizing combinability, (2) non-template methods add each atom one by one, emphasizing consistency. Semi-template methods attempt to simplify full-templates as several simpler semi-templates to achieve a balance. However, they do not consider decomposing the leaving group as a motif-tree in a controllable way. The remaining research problem is: \textit{How can we explore the full combinability-consistency space?}

\paragraph{RetroBPE} We propose a novel RetroBPE algorithm to convert the leaving groups as motif-trees. The key idea is to iteratively merge the most frequent neighboring motif pairs together into a new single motif, as shown in Algorithm~\ref{alg:motiftree} and Figure~\ref{fig:motiftree}. We introduce the time variable $t$ to describe the motif searching process: $\mathrm{M}^{(0)} \rightarrow \mathrm{M}^{(1)} \rightarrow \mathrm{M}^{(2)} \rightarrow \cdots$, where $\mathrm{M}^{(t)}$ represents the  motif set at timestamp $t$ and $\mathrm{M}^{(0)} = \{\mathcal{M}_1, \mathcal{M}_2, \cdots\}$ is initialized as a set of leaving groups whose atom number is less then $m$. Note that $\mathcal{P}_t$ records all neighboring motifs pairs at the time $t$, which serves as the candidate structural patterns to be merged. The most frequent pair are merged as the new motif $\mathcal{M}^*$. To implement the motif searching process and construct the motif tree of the leaving group $\mathcal{G}$, it is necessary to keep track of which set of atoms in $\mathcal{G}$ are grouped as motifs and then view these motifs as a basic units involved in further combinations. Once $\mathcal{M}^*$ is found, the $\mathcal{G}.\texttt{Group}(\mathcal{M}^*)$ function is used to update the motif state in the leaving group $G$. Specifically, if $\mathcal{M}^*$ can be found in $G$, the corresponding atoms are grouped as $M^*$. In Figure~\ref{fig:motiftree}, we show the process of searching motifs and mark motif pairs with dotted circles and motifs with solid circles. Finally, we emphasize several attractive features of the proposed RetroBPE:
\begin{enumerate}
  \item Treating ring structures as basic elements during the neighboring merge process to avoid breaking rings
  \item It allows users to define termination conditions based on a custom score function, which will be introduced later. 
  \item It constructs a motif tree with the point attached to the synthon molecule as the root node.
\end{enumerate}

\begin{minipage}{\linewidth}
    \centering
    \begin{minipage}{0.35\linewidth}
        \begin{figure}[H]
            \includegraphics[width=\linewidth]{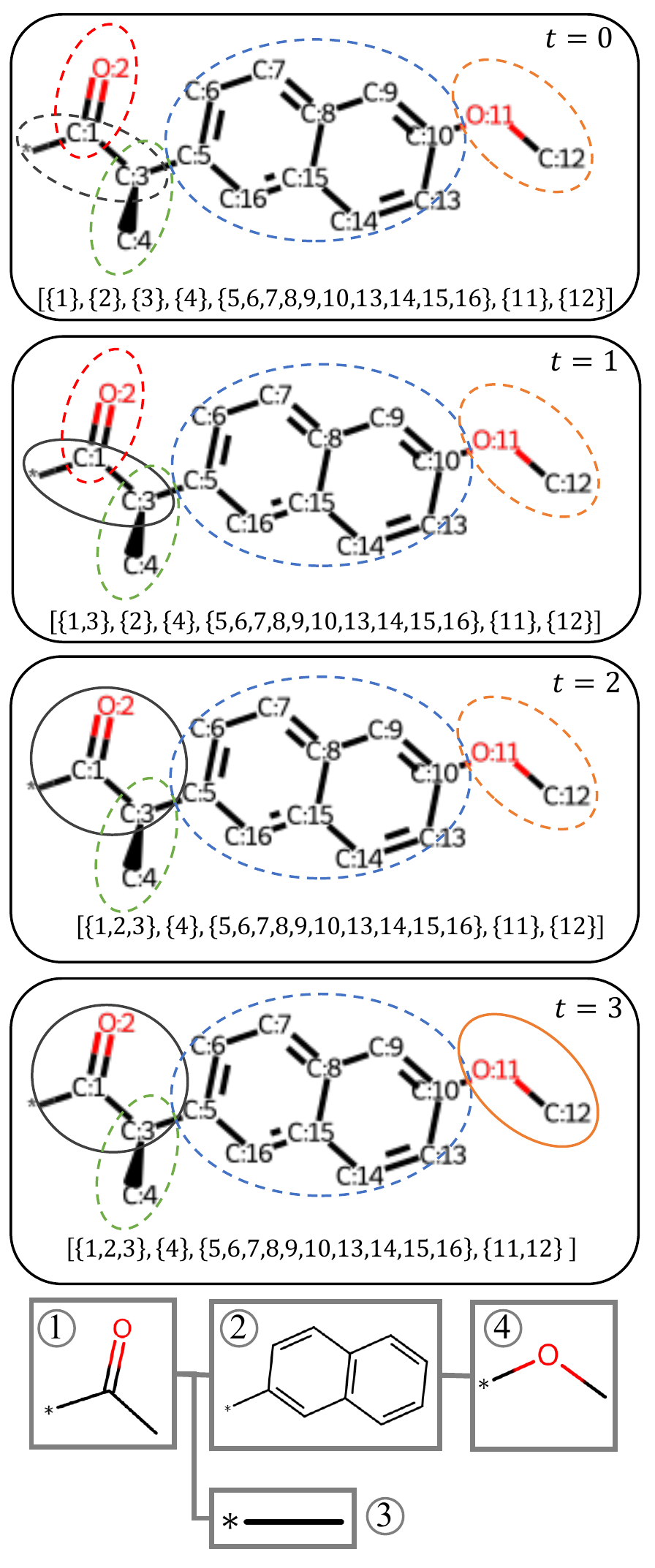}
            \caption{Motiftree construction. }
            \label{fig:motiftree}
        \end{figure}
    \end{minipage}
    \hspace{0.00\linewidth}
    \begin{minipage}{0.62\linewidth}
        \begin{algorithm}[H]
            \caption{Motif-tree construction 
            \\\textbf{Usage}: Constructing motif-tree for all leaving groups.}
            \label{alg:motiftree}
            \begin{small}
            \begin{algorithmic}[1]
                \Require Leaving groups $\mathrm{G} = \{\mathcal{G}_1, \mathcal{G}_2, \cdots\}$, the thresholds $w_{c}$ and $w_{k}$, max \#atom of the initial motif $m$, frequency function $f(\cdot)$; 
                \Ensure Motif trees $\mathcal{T} = \{\mathcal{T}_1, \mathcal{T}_2, \cdots\}$.
                \NoDo
                \NoThen
                
                \State {\color{blue} Step1:} Get initial motifs.
                \State Get initial motifs $\mathrm{M}^{(0)} = \{ \mathcal{M}_i | \mathcal{M}_i \in \mathrm{G} \quad \text{and} \quad |\mathcal{M}_i|<m \}$

                \State
                \State {\color{blue} Step2: } Iteratively merge the most frequent motif.
                \For{$t \in \mathbb{N}$}  %\Comment{Find and record new motifs.}
                    \State $\mathcal{P}_t = \{ \mathcal{M}_a \cup \mathcal{M}_b | \mathcal{M}_a \text{ and } \mathcal{M}_b \text{ are neighborhood in any } \mathcal{G}_i \}$

                    \State $\mathcal{M}^* = \argmax_{s \in \mathcal{P}_t}(f(s))$ \Comment{The most frequent motif $\mathcal{M}^*$.}
        
                    \If {$\mathcal{C}'(\mathrm{M}^{(t)})<w_{c}$ and $\mathcal{K}'(\mathrm{M}^{(t)})>w_{k}$} \Comment{Stop condition.}
                        \State Break
                    \EndIf

                    \State $\mathrm{M}^{(t+1)} = \mathrm{M}^{(t)} \cup \{ \mathcal{M}^* \}$.
        
                    \For{$\mathcal{G}_i \in \mathrm{G}$} \Comment{Update $\mathcal{M}^*$ in each leaving group.}
                        \State $\mathcal{G}_i.\text{Group}(\mathcal{M}^*)$
                        
                    \EndFor
                    
                \EndFor
        
            \State
                % Readout path
            \State {\color{blue} Step3:} Reading out motif trees for all leaving groups.
            \State $\mathcal{T} = \{\}$
            \For{$\mathcal{G}_i \in \mathrm{G}$}
                \State $\mathcal{T}_i = \text{Graph}()$ \Comment{Init graph}
                \For{$\mathcal{M} \in \text{BFS}(\mathcal{G}_i)$} \Comment{BFS or DFS traversal graph}
                    \State $\mathcal{T}_i.\text{add\_node}(\mathcal{M})$
                \EndFor
                \State $\mathcal{T} = \mathcal{T} \cup \{ \mathcal{T}_i \}$
            \EndFor
            \end{algorithmic}
        \end{small}
        
        \end{algorithm}

        \footnotetext{The most frequent neighboring motifs are iteratively merged, and the motif states are updated for each leaving group. We construct a motif tree for each leaving group, where (1) the root node of the tree being at the binding position of the leaving group to the synthons, (2) the ring being treated as the same basic unit as the atom, with no broken ring operation, and (3) each leaf node being connected to the tree through a shared atom. }
        
    \end{minipage}
\end{minipage}

% In this study, the scoring function considers two aspects: the combinability and consistency of the motifs. 

\paragraph{Scoring Function} RetroBPE employs a custom scoring function, $\text{Score}(s^*)$, to define termination conditions that meet specific requirements. In this study, the scoring function considers two aspects, including the combinability and consistency of the motifs. Let the extracted motif set be denoted as $\mathcal{M} = \{ \mathcal{M}_i | i=1,2,\cdots, N \}$, where $|\mathcal{M}_i|$ and $f(\mathcal{M}_i)$ represent the number of atoms and occurrence frequency of the $i$-th motif. The meta motif set is represented by $\dot{\mathcal{M}} = \{ \dot{\mathcal{M}}_i | i=1,2,\cdots, \dot{N} \}$, where each motif is either an atom or a ring structure. We define the combinability and consistency scores of $\mathcal{M}$ as:
\begin{equation}
    \label{eq:attention_weight}
    \begin{cases}
        \mathcal{C}(\mathcal{M}_i) = \frac{ \sum_{i=1}^{\dot{N}} f(\dot{\mathcal{M}}_i) }{\sum_{i=1}^{\dot{N}} |\dot{\mathcal{M}}_i|} \bigg/ \frac{ \sum_{i=1}^{N} f(\mathcal{M}_i) }{\sum_{i=1}^N  |\mathcal{M}_i| } & \text{\textcolor{gray}{Combinability score}} \\ % 单原子平均预测步长的比值
        \mathcal{K}(\mathcal{M}_i) = \frac{\sum_{i=1}^{N} f(\mathcal{M}_i)}{N} \bigg/ \frac{\sum_{i=1}^{\dot{N}} f(\dot{\mathcal{M}}_i)}{\dot{N}} & \text{\textcolor{gray}{Consistency score}}% component的平均出现次数比值
    \end{cases}
 \end{equation}
 where the combinability score is the ratio of the average edit lengths of adding an atom, and the consistency score is the ratio of average occurrence frequency of the motif.  A higher combinability score indicates that more atoms can be combined into a single motif, resulting in fewer editing steps needed to predict a leaving group. On the other hand, a higher consistency score indicates that the motifs occur more frequently on average, leading to greater consistency. To ensure the two scores have the similar scales, we normalize the scores between [0,1] using the following formula: $\mathcal{C}' = \mathcal{C}/\mathcal{C}_{max}$ and $\mathcal{K}' = \mathcal{K}/\mathcal{K}_{max}$, where $\mathcal{C}_{max}$ and $\mathcal{K}_{max}$ are the maximum values of the combinability and consistency scores, respectively. Finally, we stop condition can be written as:

 \begin{equation}
    \begin{cases}
        \text{False}, \text{if} \quad \mathcal{C}'(\mathcal{M}_i) < w_{c} \quad \text{and} \quad \mathcal{K}'(\mathcal{M}_i) > w_{k} \\
        \text{True}, \text{otherwise}
    \end{cases}
 \end{equation}

 We can achieve different combinability-consistency trade-offs by changing the thresholds $w_{c}$ and $w_{k}$, resulting in different algorithms. For example, MotifRetro becomes MEGAN \citep{sacha2021molecule} model if we set $(w_c, w_k)=(0,1)$, and MotifRetro becomes GraphRetro \citep{somnath2021learning} if we set $(w_c, w_k)=(1,0)$.

\subsection{Motif Editing Model}

\begin{minipage}{\linewidth}
    \begin{minipage}{0.55\linewidth}
        \paragraph{Retro-editing process}
        We model the dynamic editing process from the source molecule $\mathcal{G}^{(0)}$ to the target molecule $\mathcal{G}^{(T)}$, where four types of structural editing actions, namely \texttt{del\_bond}, \texttt{change\_atom}, \texttt{change\_bond} and \texttt{add\_motif}, will be applied to the molecular graph, leading to structural changes. We use a neural model $f_{\theta}$ to learn the dynamics of molecule editing, formulated as:
        \begin{align}
            a_t &= f_{\theta}(\mathcal{G}^{(t)}, t)\\
            \mathcal{G}^{(t+1)} &= \text{Apply}(\mathcal{G}^{(t)}, a_t)
        \end{align}
        where $\theta$ represents learnable parameters, $\mathcal{G}^{(t)}$ is current graph at the time $t$, $a_t$ is the predicted action. By applying the action $a_t$ on $\mathcal{G}^{(t)}$, we could obtain the next graph $\mathcal{G}^{t+1}$, as shown in Figure~\ref{fig:apply_action}. Note that the $\text{Apply}(\cdot)$ operation is deterministic, and the action $a_t$ needs to be predicted. We visualize different types of actions in the appendix.

    \end{minipage}
    \hspace{0.00\linewidth}
    \begin{minipage}{0.45\linewidth}
        \begin{figure}[H]
            \centering
            \includegraphics[width=\linewidth]{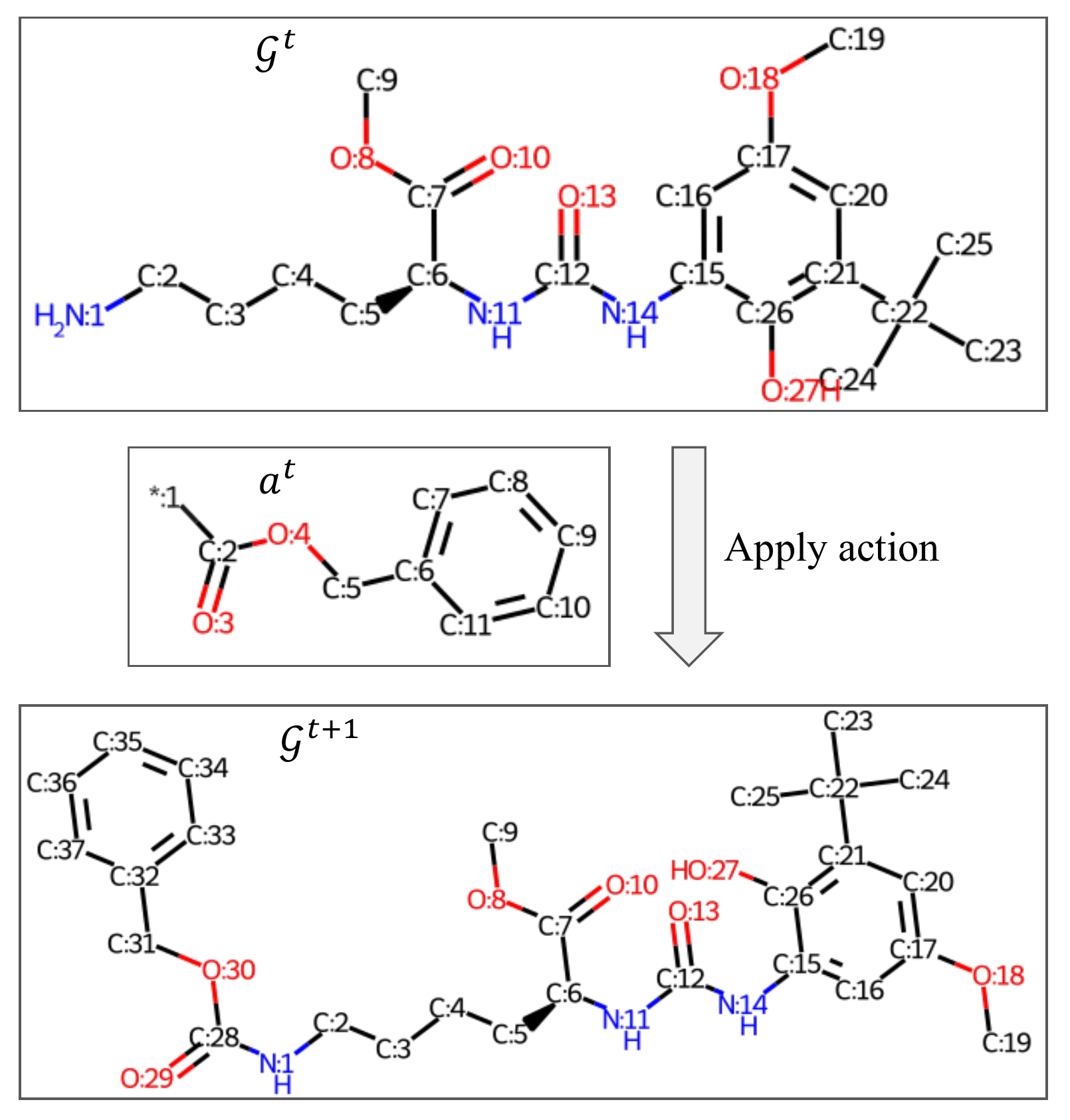}
            \caption{Apply action.  }
            \label{fig:apply_action}
        \end{figure}
    \end{minipage}
\end{minipage}

\paragraph{Dynamic Encoder} We adopt the local and global edge-updating graph attention layer (LG-EGAT) to learn molecular representations. For the input molecule $\mathcal{G}^{(t),l}(A,X^{(t),l},E^{(t),l})$, the computation process of the $l$-th encoder layer at timestamp $t$ is
\begin{equation}
    \begin{cases}
       \boldsymbol{a}_{i,j} = \frac{\exp{(\text{MLP}_1(\boldsymbol{x}_{i}^{(t),l}||\boldsymbol{x}_{j}^{(t),l}||\boldsymbol{e}_{ij}^{(t),l}))}}{\sum_{(i,j) \in A}\exp{(\text{MLP}_1(\boldsymbol{x}_{i}^{(t),l}||\boldsymbol{x}_{j}^{(t),l}||\boldsymbol{e}_{ij}^{(t),l}))}} \\
       \boldsymbol{h}_{i} = \sum_{k \in \mathcal{N}_i} a_{i,k} \text{MLP}_2(\boldsymbol{x}_k^{(t), l}) & \text{\textcolor{gray}{Local message passing attention}}\\
       \boldsymbol{g}  = \text{Pooling}(X^{(t), l}, A) \\
       \boldsymbol{x}_{i}^{(t), l+1} = \boldsymbol{h}_{i} + \text{MLP}_{3}(\boldsymbol{g}) & \text{\textcolor{gray}{Global message passing}}\\
       \boldsymbol{e}_{i,j}^{(t), l+1} = \text{EdgeMLP}(\boldsymbol{h}_{i}||\boldsymbol{h}_{j}||\boldsymbol{e}_{i,j}^{(t), l}) & \text{\textcolor{gray}{Update edge}}
    \end{cases}
\end{equation}
where $\boldsymbol{x}_{i}^l \in X^l$ and $\boldsymbol{e}_{i,j}^l \in E^l$ are node and edge features of the $l$-th layer, $\text{Pooling}(\cdot)$ is the graph pooling operation, and $||$ and $\odot$ indicate concatenation and element-wise product operation, respectively. Assuming that the encoder has $L$ layers, the final node embedding at timestamp $t$ is fed into the first layer of the network at the next timestamp as a memory state to learn the dynamics of the editing process:
\begin{align}
    \boldsymbol{m}_i^{(t),0} = \boldsymbol{x}_i^{(t-1), L}\\
    \boldsymbol{x}_i^{(t),0} \leftarrow \text{Max}(\boldsymbol{x}_i^{(t),0}, \boldsymbol{m}_i^{(t),0})
\end{align}
where $\boldsymbol{m}_i^{(t),0}$ is the initial memory state of the $i$-th node at timestamp $t$, and $\boldsymbol{m}_i^{(0),0}$ is a zero vector.

\paragraph{Decoder \& Loss} At time $t$, all the output node features and edge features are concatenated in a feature matrix $\boldsymbol{H}_{out} \in \mathbb{R}^{(|X_t|+|E_t|, d)}$, where the first $|X_t|$ feature vectors are node features and the following are $|E_t|$ edge features. These feature vectors are projected as the predictive probabilities via a linear layer:
\begin{equation}
    \boldsymbol{p} = \text{GraphSoftmax}(\text{PredMLP}(\boldsymbol{H}_{out})) \in \mathcal{R}^{(|X_t|+|E_t|)\cdot c}
\end{equation}
where $c$ is the maximum number of classes, $\text{GraphSoftmax}$ indicates that we apply \texttt{softmax} operation over all the node and edge features of the same graph. We use the binary cross entropy loss to optimize the model:
\begin{equation}
    \mathcal{L} = -\sum_{i} y_i \cdot \log{p_i} + (1-y_i) \cdot \log{(1-p_i)}
\end{equation}
Note that $y_i \in \{0,1\}$ is the ground truth label.
\vspace{-3mm}
\section{Experiments}
\vspace{-2mm}
We conduct systematic experiments to evaluate the performance of the proposed MotifRetro and will answer the following questions:
\vspace{-2mm}
\begin{itemize}[leftmargin=5.5mm]
   \item \textbf{Performance (Q1):} Could MotifRetro achieve SOTA accuracy on USPTO-50K?
   \item \textbf{Trade-off (Q2):}  How does the combinability-consistency trade-off affect the model performance?
\end{itemize}

\vspace{-2mm}
\paragraph{Datasets}  We evaluate MotifRetro on the USPTO-50k dataset\cite{schneider2016s}, which contains around 50k atom-mapped reactions with 10 reaction types. We use the same dataset split as MEGAN \citep{sacha2021molecule}, where the training/validation/test split ratio is 8:1:1.  
The original input molecules were represented as canonical SMILES during training and testing to avoid any issues with information leakage \cite{NEURIPS2020_819f46e5,somnath2021learning}. By evaluating MotifRetro on the widely used USPTO-50k dataset, we believe that we can sufficiently reveal how the combinability and consistency affect model performance.

\vspace{-2mm}
\paragraph{Implementation details}  We use the open-source cheminformatics software RDKit \cite{Landrum2016RDKit2016_09_4} to preprocess molecules and SMILES strings. The graph feature extractor consists of 6 stacked LG-EGAT, with the embedding size of 1024 for each layer. We train the proposed models 50 epochs with batch size 128 on a single NVIDIA A100 GPU, using the Adam optimizer and OneCycleLR \cite{smith2019super} scheduler. The atom and bond features can be found in the Appendix.

\vspace{-3mm}
\subsection{Performance (Q1)}
\paragraph{Objective \& Setting} We compare MotifRetro with various baselines on USPTO-50K and report the best accuracy that we can achieve. Experiments are conducted under the class-known and class-unknown settings, indicating whether or not the reaction types are fed into the model. Since multiple valid retrosynthesis results may exist, we report the top-10 accuracy using the beam search algorithm, where the beam size is 10.

\vspace{-3mm}
\begin{table}[t]
    \centering
    \caption{Overall performance. MotifRetro belongs to the semi-template or none-template category, where the best and sub-optimum results are highlighted in bold and underline. }
    \label{tab:retrosynthesis}
    \resizebox{0.8 \columnwidth}{!}{\begin{tabular}{ clcccccccccc } \hline
    \multicolumn{2}{c}{\multirow{3}{*}{\makecell{\\ \\ $k$}}}          & \multicolumn{8}{c}{top-$k$ accuracy}                                                    \\ \cline{3-10} 
    \multicolumn{2}{c}{}                             & \multicolumn{4}{c}{Reaction class known} & \multicolumn{4}{c}{Reaction class unknown} \\ \cmidrule(lr){3-6} \cmidrule(lr){7-10} 
    \multicolumn{2}{c}{}                             & 1        & 3        & 5        & 10      & 1         & 3        & 5        & 10       \\ \hline
    \multirow{5}{*}{Full-template} & RetroSim \citep{coley2017computer}       & 52.9     & 73.8     & 81.2     & 88.1    & 37.3      & 54.7     & 63.3     & 74.1     \\
                        & NeuralSym \citep{segler2017neural}     & 55.3     & 76.0     & 81.4     & 85.1    & 44.4      & 65.3     & 72.4     & 78.9     \\
                        & GLN \citep{dai2019retrosynthesis}           & {64.2}     & 79.1     & 85.2     & {90.0}    & {52.5}      & {69.0}     & {75.6}     & {83.7}     \\ 
                        & LocalRetro \citep{chen2021deep} & 63.9 & 86.8 & 92.4 & 96.3 & 53.4 & 77.5 & 85.9 & 92.4\\
                        & MHNreact \citep{seidl2021modern} & -- & -- & -- & -- & 50.5 & 73.9 & 81.0 & 87.9\\\hline
    \multirow{9}{*}{ \begin{tabular}[c]{@{}l@{}}Semi-template\\ \& None-template\end{tabular}  } 
    & GraphRetro \citep{somnath2021learning}    &  {63.9}     & {81.5}     & {85.2}     & {88.1}    & \underline{53.7}      & {68.3}     & {72.2}     & {75.5}     \\ 
                        & SCROP \citep{zheng2019predicting}         & 59.0     & 74.8     & 78.1     & 81.1    & 43.7      & 60.0     & 65.2     & 68.7     \\
                        & LV-Transformer \citep{chen2019learning} & --       & --       & --       & --      & 40.5      & 65.1     & 72.8     & 79.4     \\
                        & G2G \citep{shi2020graph}           & 61.0     & 81.3     & 86.0     & 88.7    & 48.9      & 67.6     & 72.5     & 75.5     \\
                        & RetroXpert \citep{NEURIPS2020_819f46e5}    & 62.1     & 75.8     & 78.5     & 80.9    & 50.4      & 61.1     & 62.3     & 63.4     \\ 
                        & MEGAN \citep{sacha2021molecule}   &  60.7  & {82.0}  & \underline{87.5}  & \underline{91.6}  & 48.1  & 70.7  & 78.4  & \underline{86.1}\\
                        & Dual \citep{sun2020energy} & \underline{65.7} & 81.9  & 84.7  & 85.9  & 53.6 & 70.7  & 74.6 & 77.0 \\ 
                        % & MARS \cite{} & 66.4 & 85.8 & 90.2 & 92.9 & 54.6 & 76.4 & 83.3 & 88.5\\
                        & RetroPrime \cite{wang2021retroprime} & 64.8 & 81.6 & 85.0 & 86.9 & 51.4 & 70.8 & 74.0 & 76.1\\
                        & RetroFormer \citep{wan2022retroformer} & 64.0 & \underline{82.5} & 86.7 & 90.2 & 53.2 & \underline{71.1} & \underline{76.6} & 82.1\\
                        & {\bf MotifRetro (Our)}      & \textbf{66.6}    & \textbf{85.8}    & \textbf{90.8}   & \textbf{94.5}   & \textbf{54.2} & \textbf{76.5}   &\textbf{83.4}  &\textbf{89.7} \\
                        \hline
    \end{tabular}}
\end{table}
\vspace{-3mm}

\vspace{-2mm}
\paragraph{Baselines} As to dynamic graph editing, we chose MEGAN as the strong baseline. In addition, we compare MotifRetro with various other algorithms to illustrate its benefits in a broader scope, including GLN \cite{dai2019retrosynthesis}, template-free G2G \cite{shi2020graph}, RetroXpert \cite{NEURIPS2020_819f46e5}, RetroSim \cite{coley2017computer}, NeuralSym \cite{segler2017neural}, SCROP \cite{zheng2019predicting}, LV-Transformer \cite{chen2019learning}, GraphRetro \cite{somnath2021learning}, MEGAN \cite{sacha2021molecule}, MHNreact \cite{seidl2021modern}, Dual model \cite{sun2020energy}, RetroPrime \cite{wang2021retroprime} and RetroFormer \cite{wan2022retroformer}. Note that we perfer to comparing with open-source methods, since the retrosynthesis problem is quite tricky and the potential information leakage has no way to be examined without open source code.

\vspace{-2mm}
\paragraph{Results \& Analysis} From Table~\ref{tab:retrosynthesis}, we can observe that MotifRetro consistently outperforms existing semi-template and none-template methods on all top-k accuracy metrics. This indicates that the proposed dynamic graph editing strategy could achieve higher accuracy compared to previous methods. Moreover, when compared with MEGAN, another dynamic editing method, MotifRetro achieves 5.9\% and 6.1\% performance gain in terms of top-1 accuracy on the class-known and class-unknown settings, respectively. The best top-10 accuracy also indicates that MotifRetro has better generalization ability than the baselines.

\vspace{-2mm}
\subsection{Trade-off (Q2)}
\vspace{-2mm}

\paragraph{Objective \& Setting}  By changing $w_c$ and $w_k$, we can generate different sets of motif trees, meeting different combinability-consistency trade-offs, resulting in different retrosynthesis algorithms. We evaluate these algorithms on USPTO-50K to see how the trade-off affects model performance. We run each experiment 10 times for more consistent results and report the average and standard variance of the top-k accuracy.

\vspace{-2mm}
\paragraph{Results \& Analysis} From Table~\ref{tab:tradeoff}, we conclude that: (1) the combinability-consistency affects top-k accuracy, and MotifRetro finds the good trade-off point on $(w_c, w_k)=(0.9113, 0.2777)$ and $(w_c, w_k)=(0.9254, 0.2739)$ for the class known and class unknown settings, respectively. (2) as the combinability increases, the top-1 and top-3 accuracy shows an increasing trend, indicating that the structural motifs help improve the hit rate to generate molecules similar to the reference ones. (3) the top-5 accuracy first increased and then decreased as the combinability increased, indicating a trade-off between combinability and consistency. (4) the top-10 accuracy tends to decrease as the combinability increases, indicating that composability is detrimental to generalization ability, while consistency helps model generalization. In summary, the combinability helps to increase the hit rate, the consistency helps to improve generalization ability. MEGAN and GraphRetro are special cases of MotifRetro when $(w_c, w_k)=(0,1)$ and $(w_c, w_k)=(1,0)$, respectively. However, there is a better trade-off between MEGAN and GraphRetro, where MotifRetro allow us to achieve the trade-off.

\vspace{-3mm}
\begin{table}[t]
    \centering
    \caption{Exploring the combinability-consistency trade-offs on USPTO-50K. We report the average accuracy and show its standard deviation in parentheses. The best results are highlighted in bold.}
    \label{tab:tradeoff}
    \resizebox{0.95 \columnwidth}{!}{\begin{tabular}{cccccccccc}
    \toprule
    \multicolumn{2}{c}{Config} & \multicolumn{4}{c}{Class Known} & \multicolumn{4}{c}{Class Unknown} \\ \cmidrule(lr){1-2} \cmidrule(lr){3-6} \cmidrule(lr){7-10}
    $w_c$          & $w_k$         & top-1  & top-3 & top-5 & top-10 & top-1  & top-3  & top-5  & top-10 \\ \midrule
        0.2011         &   1.0000          &  61.00(0.20)      &  79.44(0.21)     &  86.32(0.20)     &  93.63(0.18)      &  49.33(0.16)       &   69.05(0.25)     &   76.20(0.29)     &  86.80(0.26)      \\ 
        0.2468         &  0.8299           &  60.47(0.38)      &  82.61(0.39)     &   88.42(0.18)    &   93.64(0.16)     &  49.95(0.39)      &  70.04(0.19)      &  76.74(0.29)      &   85.88(0.21)     \\
        0.2763         &   0.7547          &  63.59(0.34)      &  82.70(0.22)     &  87.72(0.14)     &   92.79(0.14)     &  50.83(0.26)      &  69.61(0.25)      &   76.14(0.31)     &   85.34(0.21)     \\
        0.2853         &   0.7320          &  63.38(0.30)      &   83.52(0.17)    &  87.70(0.19)     &  93.26(0.16)      &    51.19(0.13)    &  70.70(0.32)      &   76.73(0.41)     &  85.34(0.42)      \\
        0.3880         &  0.5834           &  64.67(0.25)      &  84.14(0.17)     &  89.09(0.13)     &   94.00(0.12)     & 51.78(0.23)       & 72.07(0.26)       &    78.14(0.32)    &   86.18(0.22)     \\
        0.3953         &   0.5734          &   64.69(0.23)     &  84.70(0.22)     &  89.05(0.20)     &  93.64(0.17)      &  53.33(0.44)      &  73.00(0.21)      &  78.56(0.32)      &  85.44(0.27)     \\
        0.4349         &   0.5453          & 64.68(0.18)       & 84.40(0.27)      &  88.91(0.13)     &   93.26(0.18)     &  52.76(0.22)      & 73.77(0.24)       & 79.62(0.35)       &  86.36(0.19)     \\
        0.6536         &   0.3793          & 64.99(0.24)       &  84.94(0.14)     & \textbf{89.94(0.09)}      &  \textbf{93.89(0.10)}      &  53.00(0.29)      &   71.94(0.14)     &     77.47(0.25)   &   84.92(0.21)    \\
        0.7802         &   0.3220          &  65.91(0.27)      &  \textbf{85.54(0.15)}     &  89.26(0.14)     &  93.28(0.22)      &  53.16(0.14)      &  72.83(0.19)      &     78.67(0.32)   &   85.83(0.27)    \\
        0.7872         &   0.3201          &  65.38(0.30)      &  84.29(0.41)     &   88.72(0.33)    &  93.22(0.21)      &  53.74(0.25)      &  73.82(0.20)      &   \textbf{80.09(0.21)}     &  \textbf{86.98(0.28)}     \\
        0.8348         &   0.3022          &  66.24(0.34)      &  84.19(0.15)     &  88.70(0.16)     &  93.14(0.20)      &  53.15(0.26)      &  73.52(0.15)      &     79.43(0.30)   &   86.31(0.23)    \\
        0.8883         &   0.2845          &  65.76(0.29)      &   84.80(0.14)    &  88.87(0.17)     &  93.10(0.09)      &  53.74(0.30)      &  73.76(0.27)      &     79.24(0.18)   &  86.54(0.20)     \\
        0.9113         &   0.2777          &  \textbf{66.62(0.27)}      &  84.66(0.18)     &  88.44(0.22)     &   93.10(0.11)     &   53.39(0.28)     &  72.98(0.25)      &     78.48(0.37)   &   85.30(0.25)    \\
        0.9254         &   0.2739          &  65.90(0.08)      &  84.60(0.13)     &  89.02(0.11)     &  92.92(0.10)      &   \textbf{53.79(0.28)}     &  \textbf{73.90(0.19)}      &     79.33(0.20)   &   86.15(0.23)    \\
        0.9883         &   0.2615          &  65.56(0.18)      & 84.84(0.09)      &  88.96(0.14)     &  92.97(0.12)      &   53.33(0.24)     &  73.02(0.14)      &     78.33(0.20)   &   84.92(0.22)    \\
        1.0000         &   0.2605          &  65.60(0.20)      &  85.20(0.17)     &  89.58(0.17)     &   92.67(0.14)     &  53.37(0.22)      &  72.69(0.23)      &     78.41(0.29)   &   85.95(0.24)    \\
        \bottomrule
    \end{tabular}}
    \vspace{-5mm}
\end{table}

\section{Conclusion \& Limitation}
We propose MotifRetro, a dynamic molecular editing model for retrosynthesis equipped with a RetroBPE algorithm to explore the problem of combinability-consistency trade-off. The combinability helps to increase the hit rate, and the consistency helps to improve generalization ability. By exploring the full combinability-consistency space, MotifRetro not only achieves SOTA performance but also provides a framework for unifying existing graph-based retrosynthesis methods. One limitation is that the dynamic GNNs are computationally expensive due to the recurrent computation flow.

\bibliographystyle{plain}
\bibliography{MotifRetro}

\clearpage
\appendix
\section{Appendix}
\subsection{Technical details of molecular editing actions}
\begin{figure}[h]
  \centering
  \includegraphics[width=5.5in]{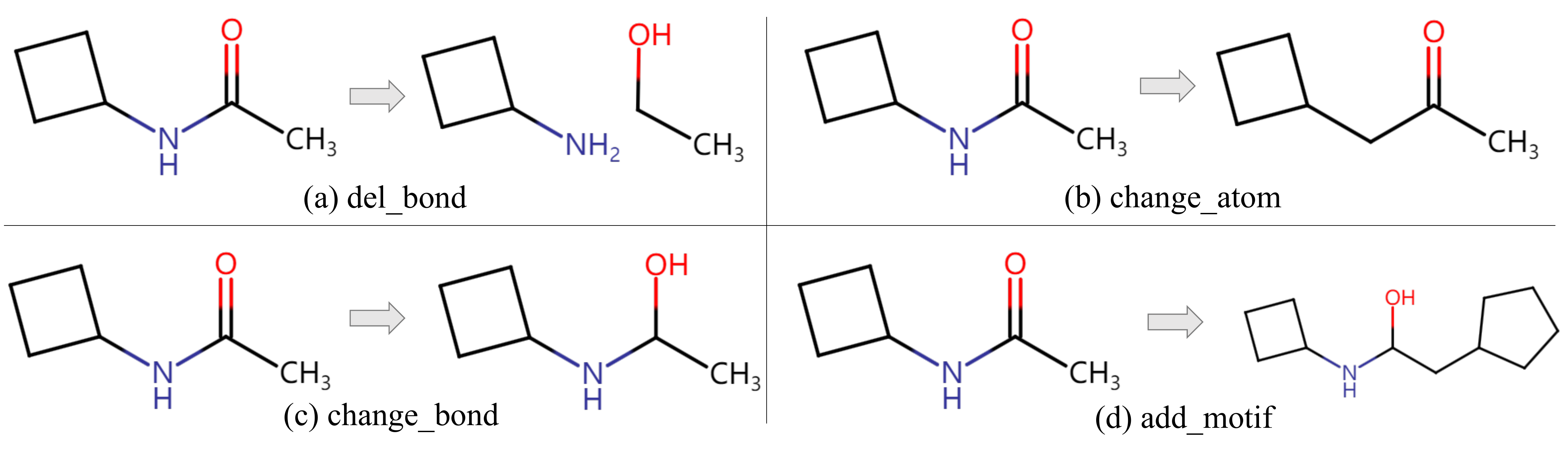}
  \caption{Visualization of actions.}
  \label{fig:action}
\end{figure}

In this subsection, we list and visualize the molecular editing actions used in MotifRetro in Table~\ref{tab:action} and Figure~\ref{fig:action}, respectively.

\begin{table}[h]
  \centering
  \caption{Molecular editing actions used in MotifRetro.}
  \label{tab:action}
  \begin{tabular}{ll}
    \toprule
  Action       & Description                 \\ \midrule
  del\_bond    & Delete bond from a molecule \\
  change\_atom & Change the atom state       \\
  change\_bond & Change the bond state       \\
  add\_motif   & Add a motif to the molecule \\ \bottomrule
  \end{tabular}
\end{table}

\subsection{Features}
In this section we list the features used in MotifRetro in Table~\ref{tab:atom_feature_center} and Table~\ref{tab:bond_feature}.

\begin{table}[h]
  \centering
  \caption{Atom features for center identification and synthon completion.}
  \label{tab:atom_feature_center}
  \begin{tabular}{cc}
  \toprule
  Name        & Description \\
  \midrule
  Atom type   & Type of atom (e.g. C, N, O), by atomic number                                     \\
  \# Hs       & one-hot embedding for the total number of Hs (explicit and implicit) on the atom \\
  Degree      & one-hot embedding for the degree of the atom in the molecule including Hs        \\
  Valence     & one-hot embedding for the total valence (explicit + implicit) of the atom        \\
  Aromaticity & Whether this atom is part of an aromatic system.                                 \\
  chiral\_tag & The chiral\_tag allows for the differentiation of stereoisomers.\\
  \bottomrule
  \end{tabular}
\end{table}

\begin{table}[H]
  \centering
  \caption{Bond features for center identification.}
  \label{tab:bond_feature}
  \begin{tabular}{cc}
  \toprule
  Name           & Description\\
  \midrule
  Bond type      & one-hot embedding for the type of the bond                 \\
  Stereo         & one-hot embedding for the stereo configuration of the bond \\
  Aromaticity & Whether this bond is part of an aromatic system.                                 \\
  \bottomrule
  \end{tabular}
\end{table}

\end{document}